\documentclass[twocolumn,showpacs,preprintnumbers,amsmath,amssymb]{revtex4}
\begin{document}

\preprint{ }

\title{ Modified Kaluza-Klein Theory, Quantum Hidden Variables and 3-Dimensional Time }  
\author{\normalsize Xiaodong Chen}
\altaffiliation{2073B Vestavia Park Ct, Birmingham, AL 35216, U.S.A.}
\email{xiaodong.chen@gmail.com} 

\date{\today}

\begin{abstract} 
In this paper,  the basic quantum field equations of free particle with 0-spin, 
1-spin (for cases of massless and mass $>$ 0) and $\frac{1}{2}$ spin are derived from Einstein equations under 
modified Kaluza-Klein metric, it shows that the 
equations of quantum fields can be interpreted as pure geometry properties of curved higher-dimensional time-space . 
One will find that if we interpret the 5th and 6th dimension as ``extra'' time dimension, the particle's 
wave-function can be naturally interpreted as a single particle moving along geodesic path in 6-dimensional  
modified Kaluza-Klein time-space. As the result, the fundamental physical effect of quantum theory such as 
double-slit interference of single particle, statistical effect of wave-function, wave-packet collapse, spin, 
Bose-Einstein condensation, Pauli exclusive principle can be interpreted as ``classical'' behavior in new time-space.
In the last part of this paper, we will coupling field equations of 0-spin, 1-spin and $\frac{1}{2}$-spin particles 
with gravity equations. 
\end{abstract}

\maketitle

\section{Introduction} \label{INTRO}

Kaluza-Klein's theory\cite{Kal21} showed that five-dimensional general relativity contains both Einstein's 4-dimensional 
theory of gravity and Maxwell's theory of electromagnetism[2]. Kaluza gave physicist the hope of unifying matter 
and geometry. Currently, people extended Kaluza-Klein's idea to a possible ``theory of everything,''
10-dimensional superstrings. 
One of main difficulty of the original Kaluza-Klein's theory was: trying to make the theory to fit on our macrocosm 
observation that why had no 
fifth dimension been observed in nature? This cause the later efforts of different versions of Kaluza-Klein's 
theory such as compactified, projective and noncompactified Kaluza-Klein theory.

In another totally different area, Bohm \cite{bohm} and etc. tried to find quantum hidden variables to interpret quantum physics
under the language of ``classical'' physics. There are many difficulties Quantum Hidden Variable (QHV) theory had to solve.
One of main issue is: in quantum physics, even a single particle can show non-local effect, but there is no such thing 
in classical physics or relativity theory. For instance, to explain the interference pattern 
in double-slit interference experiment, one has to 
accept the fact that: a single photon (or electron) has to pass both slits at the same time. In general, in quantum physics, 
a single particle can spread out in large area -- occupy many different spatial locations at the same time. 

In 1999, X.Chen \cite{xchen} proposed that using extra time dimension, we can explain why a single particle shows at two different
locations at the same time. The paper didn't derive any basic quantum equations.  In this paper, single free-particle equation for
0-spin, 1-spin (for massless and mass $>$ 0) and $\frac{1}{2}$ spin will be derived by modified Kaluza-Klein equation. Thus, we shows that the equations
of quantum fields can be interpreted as pure geometry properties of curved higher-dimensional time-space. If we assume that the 
5th and 6th ``Kaluza like'' dimensions are time dimensions, then the physical effects of these extra dimensions will show basic 
behavior of quantum particles. Wave-function of single particle becomes geodesic path in 6-dimensional modified Kaluza-Klein time-space. 
In section (\ref{ZeroSpin}), 0-spin single free particle equation (Klein-Gordon
equation) with mass $>$ 0 is derived through modified Kaluza-Klein metric. Section (\ref{Interpret}) will discuss the detail of using the two extra time dimensions to interpret 
basic quantum effects such as double-slit interference experiment, statistical effect of wave-function, Bose-Einstein condensation, 
Pauli exclusive principle.
In section (\ref{OneSpin}), Maxwell's theory of electromagnetism will be re-derived from modified Kaluza-Klein equations. 
The equation of 1-spin's free single particle with mass $m_0 >0$ will also be derived, the mass part of U(1) gauge field 
will be naturally included as derivative of 6th dimension.
In section (\ref{halfSpin}) , we will obtain Dirac field equation of single particle with $\frac{1}{2} $ spin through 6-dimensional Einstein equations. 
In section (\ref{gravity}), we will coupling field equations of 0-spin, 1-spin and $\frac{1}{2}$-spin particles 
with gravity equations.

\section{Equations of 0-spin free particle} \label{ZeroSpin}

The original Kaluza metric can be written as follows \cite{Overduin}: 
\begin{equation}
\left( \hat{g}_{AB} \right) = \left( \begin{array}{cc}
   g_{\alpha\beta} + \kappa^2 \phi^2 A_{\alpha} A_{\beta} \; \; & \; \; 
     \kappa \phi^2 A_{\alpha} \\
   \kappa \phi^2 A_{\beta} \; \;                                & \; \; 
      \phi^2
   \end{array} \right) \; \; \; ,
\label{5dMetric}
\end{equation}
where the $\alpha\beta$-part of $\hat{g}_{AB}$ with $g_{\alpha\beta}$ 
(the four-dimensional metric tensor), the $\alpha 4$-part with $A_{\alpha}$ (the electromagnetic potential),
and the $44$-part with $\phi$  (a scalar field). The four-dimensional metric signature is taken 
to be $(+ \, - \, - \, -)$

In this section, we'll only focus on obtaining field equation of 0-spin free particle, 
so we ignore vector field $A_\alpha$,  
(equation with 1-spin particle will be discussed in
section (\ref{OneSpin})). Furthermore, We will add one more extra dimension than original Kaluza metric, 
i.e. totally 6-dimensional time-space with conditions: 
\begin{eqnarray}
\partial_{4} g_{44} = 0  \;, \; \;  \;
\partial_{5} g_{44} =  -ia_{5}g_{44} \;, \; \;  \;
g_{55} =  -1   
\label{6DCondition} 
\end{eqnarray}
where $a_5$ is constant. For free particle, we ignore gravity field, then 
$g_{\alpha\beta} = {\delta}_{\alpha\beta}$, 
6-dimensional time-space metric becomes:
\begin{equation}
\left( \hat{g}_{AB} \right) = \left( \begin{array}{cc}
   g_{\alpha\beta} \; \; \; \; \; \;  \; \; \; \; \; \; \\
  \; \; \; \; \; \; \; \; g_{44} \; \; \; \; \; \; \; \; \\
   \; \; \; \;  \; \; \; \; \; \; \; \; \; \; -1 \\ 
   \end{array} \right) 
\label{5dMetric_0}
\end{equation}

The 6-dimensional Ricci tensor and Christoffel symbols are defined
in terms of the metric exactly as in four dimensions:
\begin{eqnarray}
\hat{R}_{AB}        & = & \partial_C \hat{\Gamma}^C_{AB} -
                          \partial_B \hat{\Gamma}^C_{AC} +
                          \hat{\Gamma}^C_{AB} \hat{\Gamma}^D_{CD} -
                          \hat{\Gamma}^C_{AD} \hat{\Gamma}^D_{BC} 
                          \; \; \; , \nonumber \\
\hat{\Gamma}^C_{AB} & = & \frac{1}{2} \hat{g}^{CD} \left( 
                          \partial_A \hat{g}_{DB} +
                          \partial_B \hat{g}_{DA} -
                          \partial_D \hat{g}_{AB} \right) \; \; \; .
\label{5dChristRicci}
\end{eqnarray}
where A, B.. run over 0,1,2,3,4,5. 

The 6-dimensional Einstein equations are  
\begin{equation}
\hat{G}_{AB} = \kappa T_{AB} \; \; \; ,
\label{5dEFE1}
\end{equation}
where $T^{AB}$ is 6-dimensional energy momentum tensor, 
$\hat{G}_{AB} \equiv \hat{R}_{AB} - \hat{R} \, \hat{g}_{AB} / 2$
is the Einstein tensor, $\hat{R}_{AB}$ and
$\hat{R} = \hat{g}_{AB} \hat{R}^{AB}$ are 
the 6-dimensional Ricci tensor and scalar respectively,
and $\hat{g}_{AB}$ is the 6-dimensional metric tensor, A, B.. run over 0,1,2,3,4,5 .

Using metric (\ref{5dMetric_0}), we get following non-zero Christoffel symbol:
\begin{eqnarray}
\hat{\Gamma}^{4}_{4 \alpha } = \frac{1}{2} g^{44} \partial_{\alpha} g_{44} \;, \; \; \; \; \; \; \;
\hat{\Gamma}^{4}_{{\alpha} 4} =  \hat{\Gamma}^{4}_{4 \alpha } \nonumber \\
\hat{\Gamma}^{\alpha}_{44}  =  -\frac{1}{2}g^{\alpha\alpha}\partial_{\alpha}g_{44}  \;,  \; \; \; 
\hat{\Gamma}^{4}_{45} =  \frac{1}{2}g^{44}\partial_{5}g_{44} \nonumber \\
\hat{\Gamma}^4_{54}  =  \hat{\Gamma}^{4}_{45}  \;,  \; \; \; \;
\hat{\Gamma}^{5}_{44}  =  -\frac{1}{2}g^{55}\partial_{5}g_{44} 
\label{5dChristoffel}
\end{eqnarray}
Start from here, through out this paper, capital Latin indices A,B,C .. run over 0,1,2,3,5 (A,B $<> 4$),
Greek indices $\alpha, \beta$ ... run over 0, 1, 2, 3, and small Latin indices a, b, ... run over 1, 2, 3.

Let
\begin{eqnarray}
g_{44} = (\phi(x_{0},x_{1},x_{2},x_{3})e^{-i(a_{5}x_{5})})^2 \;  \nonumber \\ 
g^{44} = (\phi^{\star}(x_{0},x_{1},x_{2},x_{3})e^{i(a_{5}x_{5})})^2\; 
\label{phi_0}
\end{eqnarray}
where $\phi^{\star}$ is complex conjugation of $\phi$,
and 
\begin{equation}
\phi = e^{-i(a_{0}x_{0}-a_{1}x_{1}-a_{2}x_{2}-a_{3}x_{3})}
\label{phi_1}
\end{equation}
where $a_{\alpha}$ is constant. 

Substitute (\ref{phi_0}) and (\ref{5dChristoffel}) in (\ref{5dChristRicci}): 
\begin{eqnarray}
R_{\alpha\beta} =  -(\phi^{\star}\partial_{\alpha}\phi) (\phi^{\star}\partial_{\beta}\phi) \nonumber \\
R_{\alpha 5} = R_{5\alpha} =  -(\phi^{\star}\partial_{\alpha}\phi) (\phi^{\star}\partial_{5}\phi)\nonumber \\
R_{55} = -(\phi^{\star}\partial_{5}\phi) (\phi^{\star}\partial_{5}\phi) \nonumber \\
R_{\alpha4} = R_{4\alpha} = 0 \nonumber \\
R_{44}  = -\phi(\partial^{\alpha} \partial_{\alpha}\phi)-\phi(\partial^{5} \partial_{5}\phi) 
\label{5dRicci}
\end{eqnarray}
and Ricci scalar becomes
\begin{equation}
R = g^{AB} R_{AB} = 0
\label{RicciScalar} 
\end{equation}
Here we used equation (\ref{phi_0}).

Using (\ref{5dRicci}),(\ref{phi_0}), and let $a_{5} = -\frac{m_0}{\hbar} $ where $m_0$ is rest mass of particle,
$\hbar$ is Planck constant,  
Einstein equations (\ref{5dEFE1}) become:
\begin{equation}
 -(\phi^{\star}\partial_{\alpha}\phi) (\phi^{\star}\partial_{\beta}\phi) = \kappa T_{\alpha\beta}
\label{5dMSTensor_0}
\end{equation}
\begin{equation}
i \frac{m_0}{\hbar}\phi^{\star}\partial_{\alpha}\phi = \kappa T_{5\alpha} =\kappa T_{\alpha 5}
\end{equation}
\begin{equation}
(\frac{m_0}{\hbar})^2 = \kappa T_{55}
\end{equation}
\begin{equation}
 \kappa T^{4 \beta} = \kappa T_{\alpha 4} = 0
\label{5dSETensor_1}
\end{equation}
\begin{equation}
-\frac{1}{\hbar^2} \partial^{\alpha} \partial_{\alpha} \phi - (\frac{m_0}{\hbar})^2\phi = 0
\label{5dSETensor_2}
\end{equation}
Here we let $T_{44} = 0$, i.e. no 5-dimensional energy momentum tensor. Equation (\ref{5dSETensor_2}) 
is Klein-Gordon Equation for free 0-spin particle. Actually, it is reasonable 
to let $T_{\alpha\beta} = p_{\alpha}p_{\beta} $, $p_{\alpha}$ is momentum vector of particle, $\alpha$ run over (0,1,2,3), 
the solution of equations (\ref{5dMSTensor_0})--(\ref{5dSETensor_2}) is:
\begin{equation}
\phi = e^{-i\sqrt{\kappa} (p^{\alpha}x_{\alpha})}
\label{Function_spin0}
\end{equation}
$a_{\alpha}$ in (\ref{phi_1}) becomes $\frac{p^{\alpha}}{\hbar} $. 

If we let $\kappa = -(\frac{1}{\hbar})^2 $, equation (\ref{Function_spin0}) become plane
wave-function of a single particle. We can see that if we describe quantum field in {\it pure geometry},
$\frac{i}{\hbar}$ plays the similar role as $\sqrt{8\pi G}$ 's 
role in 4-dimensional  Einstein equation, where G is gravational constant. The metric tensor becomes
\begin{equation}
\left( \hat{g}_{AB} \right) = \left( \begin{array}{cc}
   g_{\alpha\beta} \; \;  \; \; \; \; \; \; \; \; \; \; 
       \; \; \; \; \; \; \; \; \; \; \; \; \; \; \\
   \; \; \; \;  \; \; \; \; \; \; \; \;
    e^{-\frac{2i}{\hbar}(p^{\alpha}x_{\alpha}-m_{0}x_5)} \; \; \; \; \; \; \\
   \; \; \; \; \; \; \; \; \; \; \; \; \; \; \; \; \; \; \; \; \; \; \; \; \; \; \; \;   \; \; \; \; \; \; \; \;  -1 
   \end{array} \right) 
\label{5dMetric_1}
\end{equation}

The interval 
\begin{equation}
ds^2 = dx_0^2-dx_1^2-dx_2^2-dx_3^2+ e^{-\frac{2i}{\hbar}(p^{\alpha}x_{\alpha}-m_{0} x_5)} dx_4^2 - dx_5^2 
\label{interval}
\end{equation}
$ds^2$ becomes complex function, and in this paper, we always choose $c \equiv 1$.
The length can be define as
\begin{equation}
dl = \sqrt{\left | ds \right |^2}
\label{length}
\end{equation}
where $|ds|$ is mod of $ds$.

\section{A New Interpretation -- Time as 5th and 6th dimensions} \label{Interpret}

\subsection{Time as quantum hidden variable} \label{Interpret_0}

Before we discuss the meaning of metric (\ref{5dMetric_1}), let's go back to basic quantum physics. 
In double-slit 
interference experiment of photons, it is well known that to get interference pattern, one has to 
assume that 
both slits
affect each single photon, even if we make light beam weak enough to emit almost just 1 photon each time, 
we will still 
get interference fringes, but if we tried to measure which slit photon passed, the interference 
fringes will be destroyed.
If one trys to use path to describe the movement of a single photon, he has to say that: a single photon 
passes two slits at 
the same time !  It sounds against our common physics law, that's why in current quantum physics, 
we have to say that 
there is no path for quantum particle, even in Fenyman's path integral, we have to interpret 
that the path is imaginary 
path, it is not the real path of particle's movement. Similar non-localized behavior can be found in 
Bose-Einstein condensation and 
Superconductivity theory, one has to assume that the wave function of each single particles spread out the whole lattice
--- each single particles is everywhere in lattice, it is so called identity particles.  Should we satisfy the answer 
from traditional quantum physics: there is no path in quantum world? Or there are some mysterious paths which 
we haven't
understood yet? Quantum hidden variable theory tried to find the hidden variables so that we can describe the 
quantum effect
by using classical physics language (including ``path'' concept). No quantum hidden variable theory gives us a satisfied answer
yet.

Let's re-phrase the statement we discussed: {\it To use path to describe a single particle's motion, we have to say: 1 particle 
can occupied 2 (or more) spatial locations at the same time}. It is against our common sense -- how can a particle shows up in
two locations at the same time and it is still the same particle ? If you see Bob at Las Vegas at 9:00 am Jan. 1st, 2004
 central time, your brother see Bob at New York at 9:00 am Jan. 1st 2004 central time, how can you tell that you and your brother
 are seeing the same Bob, not Bob's twin brother? The way to find out is: if you break Bob's left hand one minute later(at 9:01 am) in Las Vegas 
 (as an extreme way to affect the physical state of the object), then if your brother see that Bob's left hand suddenly broken at 9:01 am in New York,
then he is the same Bob (his physical state affected by your behavior), but if the Bob who your brother seeing is still with a perfect left hand, then he must
 be Bob's twin brother, not Bob himself. Suppose that your brother does see Bob's left hand suddenly broken in New York, then there must be something wrong in our time-space. 
 First we are sure you and your brother are seeing the same Bob (broken left hand), then you and your brother are seeing Bob at different spatial locations
($x_1,x_2,x_3$)(Las Vegas and New York), then you and your brother are seeing Bob at the same time t -- Bob can not travel faster than light, so we 
have way to ensure that time is the same (ignore the gravity of the earth). How could we interpret this? Let us ask a question, is the same time t means the same time?
If time is more than 1 dimension, there is another hidden dimension of time $t_{\theta}$, so that Bob in New York at time $(t,t_{\theta2})$,
Bob in Las Vegas at time $(t,t_{\theta1})$. Bob uses $t_{\theta2}-t_{\theta1}$ traveling from Las Vegas to New York and then travel back and 
forth, but we don't have any apparatus to measure $t_{\theta}$, also sine it is $t_{\theta}$, not t, we can not use speed to measure the travel from to spatial 
location by $t_{\theta}$. If Bob accidentally dead in Las Vegas at 9:02 am, he can not travel to New York through second time dimension $t_{\theta}$, so he will disappear in  
New York at 9:02 am, which is called wave-package collapse in quantum physics. 

To understand the above discussion better, we need understand what will happen if there are extra time dimensions. 
In our 1-dimensional time world, time tells us the order of events happened. We are doing things always in the time order. We can not travel to
two different places at the same time because of 1 dimensional time; Time is the only parameter to describe the motion of object. 
If there is an extra time dimension, that means the order of a serial events happened for an object will be described by two parameters 
Then event A and event B which happed at the same time t in 1 dimensional time world, could actually happened at two different times in
2-dimensional time world -- A and B at the same time t1, but different time t2. i.e. Serial events in 2 dimensional time, 
could be seen as parallel events if you only use 1 dimension clock. 
The new time t2 is hidden order of even happening in our 1-dimensional time world.

Imagine if we only have 1 dimensional knowledge of the world but the actually world is 3-dimension, so we use ruler to measure 
space and we do not have concept of direction, then if we pick up two points on a circle with angle $\theta$ ,
($ 0 < \theta < \frac{\pi}{2} $) and our the measurement start from center of circle, by using 1-dimensional language,
we will think that the two points are the same point since the distance is the same. 

Extending the above discussions about Bob to quantum world: a single particle can occupy more than one spatial locations at the 
time t, if we change the state of 
the particle in one location $X(x_1,x_2,x_3)$ at time t, its state 
in another location $Y(y_1,y_2,y_3)$ will be changed at the same time t. If we localized the particles location at  
$X(x_1,x_2,x_3)$ at time t, the particle can not be shown at location $Y(y_1,y_2,y_3)$ at time t anymore. {\it These are the real phenomena 
observed in quantum world!} It is only valid in quantum world, since wave-length
$\lambda = \frac{h}{p} $, where p is momentum of particle, in macrocosm world, p is too big and $\lambda $ is too small.
To interpret this by using language of ``classical'' physics, 
we need two hypotheses:

1) There is at least one extra time dimension in our world, the new time dimension (or dimensions) acts like Kaluza's 
5th dimension, it is a loop wrapped around the rest 4-dimension time-space. 

2) A particle's motion is determined by its local curved geometry properties of multiple time dimensions + 3 dimensional spaces;
A particle moves along geodesic path in its local curved time-space. 

If the first hypotheses is true, then ``a particle shows in more than one locations'' at the same time t is because of extra time dimensions.
The particle moves from one location to another location at the same time t by extra time dimension $t_2$. Later, we will prove that ``the path 
of particle's motion by extra time dimension'' is plan-wave and it is geodesic in 6-dimensional time-space. 

Should we assume the extra time dimension is a ``small'' loop? It is not necessary. If the particle always stays in the same location at time t in 
all 2nd and 3rd time dimensions $t_{2}, t_{3}$, then the particle will behave the same as an object in classical physics, so even if 2nd or 3rd time
dimension is big, as long as the whole loop of extra dimensional time $t_{2}$ and $t_{3}$ always stay in one location at each first dimensional time t, 
we will not see any effect of extra dimensional time. The curvature of local time-space is in quantum level -- ``small''.
The reason we can not see extra time dimension in macrocosm
is because: all objects' motion in our macrocosm world are collective motion of enormous particles, the effect of the extra time dimension of each particles counterpart to
each other. In addition,  in Einstein theory, gravity potential is caused by curved time-space, we 
extend Einstein's idea: a particle's energy makes its local space curved. As we all know that, even in non-relativity quantum physics, a particle's energy is always $>$ 0. If 
the above two hypotheses are correct, then a particle's local time-space is always curved: no curved time-space, no quantum effect of particle, then particle can not exist. 
Based on two hypotheses, we can start interpret quantum phenomena.
 
\subsection{Geodesic of 6-dimensional time-space} \label{Interpret_1}

Now we are ready to use the results in section \ref{ZeroSpin}. It gives us the 6-dimensional metric and equations of 
0-spin particle's local time-space. By using the results in section \ref{ZeroSpin}, we are able to show that wave-function of particle 
is geodesic in 6-dimensional time-space.

The 6-dimensional geodesic equation is :
\begin{equation}
\frac{d^{2}x^A}{d\tau^2} + \Gamma^{A}_{BC} \frac{dx^B}{d\tau}\frac{dx^C}{d\tau} = 0
\label{Geodesic}
\end{equation}
where $\tau$ is affine parameter. Using equation (\ref{5dChristoffel} ), above equation becomes:
\begin{eqnarray}
\frac{d^{2}x^{\alpha}}{d\tau^2} + \Gamma^{\alpha}_{44} \frac{dx^4}{d\tau}\frac{dx^4}{d\tau} = 0 \\
\frac{d^{2}x^4}{d\tau^2} + 2\Gamma^{4}_{4\alpha} \frac{dx^4}{d\tau}\frac{dx^{\alpha}}{d\tau} = 0 
\label{Geodesic_1}
\end{eqnarray}
where $\alpha$ run over 0,1,2,3,5.

To find the solution for (\ref{Geodesic}), 
we need find the relations between $x^A$ and $\tau$.  let:
\begin{equation}
x^{\alpha} = \frac{-i p^{\alpha}}{2\hbar}\tau^2 + constant
\label{Space_Time_Relation_1}
\end{equation}
where $\alpha$ runs over 0,1,2,3.
\begin{equation}
x^{5} = \frac{-i m_{0}}{2\hbar}\tau^2  + constant
\label{Space_Time_Relation_2}
\end{equation}
and
\begin{equation}
x^{4} = e^{\frac{i}{\hbar}(p^{\alpha}x_{\alpha} - m_{0} x_5)}\tau + constant 
\label{Space_Time_Relation_3}
\end{equation}
the constants in (\ref{Space_Time_Relation_1}), (\ref{Space_Time_Relation_2}), (\ref{Space_Time_Relation_3}) are determined 
by initial value of $x^{A}$. Then:
\begin{eqnarray}
\frac{dx^4}{d\tau} = e^{\frac{i}{\hbar}(p^{\alpha}x_{\alpha}-m_{0}x_5)}(1 + 
                 \frac{i}{\hbar}(p^{\alpha}\frac{dx^{\alpha}}{d\tau} - m_0\frac{dx^5}{d\tau})\tau) \nonumber \\
                  = e^{\frac{i}{\hbar}(p^{\alpha}x_{\alpha}-m_{0}x_5)}
		    (1+(\frac{i}{\hbar})^2(p_0^2-p_a^2- m_0^2)\tau^2) \nonumber \\     
\label{x4geodesic_0}
\end{eqnarray}
where $a$ runs over 1,2,3. After use the relation: $p_0^2 = p_1^2+p_2^2+p_3^2+m_0^2$, equation (\ref{x4geodesic_0}) becomes:
\begin{equation}
\frac{dx^4}{d\tau} = e^{\frac{i}{\hbar}(p^{\alpha}x_{\alpha}-m_{0}{\hbar}x_5)}
\label{x4geodesic}
\end{equation}
Using similar algebra, we can get:
\begin{equation}
\frac{d^2x^4}{d\tau^2} = 0
\end{equation}
It is easy to examine that, (\ref{Space_Time_Relation_1}) 
(\ref{Space_Time_Relation_2}) (\ref{Space_Time_Relation_3}) are the solutions of (\ref{Geodesic}).
i.e. $\tau$ is affine parameter of geodesic and the equations above are solutions of geodesic.
Rewrite (\ref{Space_Time_Relation_1}):
\begin{equation}
\frac{dx^{\alpha}}{d\tau} = \frac{-i p^{\alpha}}{\hbar}\tau 
\label{Space_Time_Relation_4}
\end{equation}
Rewrite (\ref{Space_Time_Relation_3}) and (\ref{x4geodesic_0}):
\begin{eqnarray}
\tau = e^{\frac{-i}{\hbar}(p^{\alpha}x_{\alpha} - m_{0} x_5)} x^4  \\
\frac{d\tau}{dx^4} = e^{\frac{-i}{\hbar}(p^{\alpha}x_{\alpha} - m_{0} x_5)} 
\label{Space_Time_Relation_5}
\end{eqnarray}
Then using (\ref{Space_Time_Relation_4}), (\ref{Space_Time_Relation_5}) :
\begin{equation}
\frac{dx^{\alpha}}{dx^4} = \frac{dx^{\alpha}}{d\tau} \frac{d\tau}{dx^4}
=\frac{-i p^{\alpha}}{\hbar}e^{\frac{-2i}{\hbar}(p^{\alpha}x_{\alpha} - m_{0} x_5)} x^4
\label{Space_Time_Relation_6}
\end{equation}

Similarly:
\begin{eqnarray}
\frac{dx^5}{dx^4} = \frac{i m_0}{\hbar} 
           e^{-\frac{2i}{\hbar}(p^{\alpha} x_{\alpha} - m_{0}x_5)} x^4  
\label{Space_Time_Relation_7}
\end{eqnarray}
Thus:
\begin{eqnarray}
ds = \sqrt{1- \frac{(g_{\alpha \alpha} p^{\alpha}p^{\alpha} - m_0^2)}{\hbar^2} e^{\frac{-4i}{\hbar} p^{A} x_{A}} x_4^2 }\, e^{-\frac{i}{\hbar} p^{A} x_{A}} dx_4 \nonumber \\
  = e^{\frac{-i}{\hbar}(p^{\alpha} x_{\alpha}- m_0 x_5)} dx_4 \nonumber \\
\label{Geodesic_2}
\end{eqnarray}
where A,B runs over 0,1,2,3,5, and $\alpha$ runs over 0,1,2,3 and $p^5 = -m_0$,
so 6-dimensional geodesic becomes plane wave-function by $dx_4$. In addition, from 
(\ref{Space_Time_Relation_1}), one can see:
\begin{equation}
\frac{dx^{\alpha}}{dt} = \frac{p^{\alpha}}{p^0} = v
\end{equation}
where we let speed of light c = 1, v is ``classical'' speed of particle.

Re-write equation (\ref{Space_Time_Relation_6}) as:
\begin{equation}
\triangle x_{\alpha} = \frac{i p^{\alpha}}{2\hbar}e^{\frac{-2i}{\hbar} p^{A}x_{A}} (g^{44})^2 \triangle (x_4)^2  
   = \frac{i p^{\alpha}}{2\hbar}e^{\frac{2i}{\hbar} p^{A}x_{A}} \triangle (x_4)^2
\label{Superstition_0}
\end{equation}
It gives us
\begin{eqnarray}
|\triangle (x_4)^2|  = |\frac{-2i\hbar}{ p^{\alpha}}e^{\frac{-2i}{\hbar} p^{A}x_{A}}| |\triangle x_{\alpha}| \nonumber \\
   = |\frac{-2i\hbar}{ p^{\alpha}}\psi^2| |\triangle x_{\alpha}|
\label{Superstition_1}
\end{eqnarray}
where $\psi = e^{\frac{-i }{\hbar} (p^{\alpha}x_{\alpha} - m_{0} x_5)} $ is wave-function. 
We need this equation in next sub-section.

Finally, from (\ref{Space_Time_Relation_3}), one can see that: in vaccum, without any particle, $p_{A}=0$, then $dx_4 = d\tau$.    
In relativity, proper time $\tau_p$ is affine parameter, time t has relationship with $\tau_p$:  $d\tau_p = \sqrt{(1-v^2)} dt $. 
Time t becomes $\tau_p$ in special frame that $v=0$. Now we see similarity between $x_4$ and $x_0$ -- under certain condition,
they both become affine parameter. Equation (\ref{Geodesic_2}) tells us that $x_4$ alone can not keep $ds$ at fixed value,
when $x_4$ unchange, we still can change other $x_{0}$ and $x_{5}$ to make particle's $ds$ change (the other three spactial coordinate
$dx_1$, $dx_2$, $dx_3$ should depend on time coordinates), so we have three parameters
to describe particle's motion (or ``order'' of events) -- 3 dimensional time.

\subsection{Quantum effects of 6-dimensional time-space} \label{Interpret_2}

Before we go further, we need discuss the measurement of particle. 

How do we measure a particle in 6-dimensional time-space? We use an apparatus to measure a particle P, first the 
apparatus need meet particle P, that means at least one particle A of apparatus must meet the particle P at
some points of 6-dimensional time-space; Or in other words, particle A's geodesic and particle P's geodesic must cross 
each other at some 6-dimensional points. At those points, the 6-dimension coordinates of particle P equals to 6-dimensional coordinates of
particle A. If at spatial location $X(x_1,x_2,x_3)$ at time t (through this paper, we always use X as 3-space dimension and t 
as first time dimension) , the 2nd and 3rd time coordinate of particle P is $(x_{4P},x_{5P})$, 
and the 2nd and 3rd time coordinate of particle A is $(x_{4A},x_{5A})$, if $x_{4P} \neq x_{4A} $ or $x_{5P} \neq x_{5A}$, then even if
P and A
both show at location X at the same time t, they do not meet (they are different points in 6-dimensional time-space), so A and P can not ``see'' each other.
Therefore, the chance of find P at (X,t) is not determined, it is rather a statistical result.

As we discussed before, P's geodesic is not a straight line, if it passes (X,t) more than once through 2nd and 3rd time dimensions,  
then particle A has better chances to meet P. The chances of meeting P at (X,t) is depended on how many $(x_{4P},x_{5P})$ of particle P ``pass'' (X,t).
i.e., the possibility of A meet P at (X,t) is proportional to the density of $(x_{4P},x_{5P})$ pass through (X,t). 
From equation (\ref{Superstition_1}), in a small area $\triangle x_{\alpha}$, the density $|\triangle (x_4)^2|$ is
proportional to $|\psi^2|$, since the possibility of finding P in $\triangle x_{\alpha}$ is proportional to density of $x_4$,
then the possiblity of finding P in $\triangle x_{\alpha}$ is proportional to $|\psi^2|$. 
Now we get the same conclusion as the statistical interpretation of wave-function! We don't need 
considering the density of $x_5$ because: we chose a special coordinate system to make metric equals (\ref{5dMetric_0}), 
under this metric, the geodesic only depends on $x_4$ as we saw in (\ref{Geodesic_2}).

In double slits experiment, a particles path is
splited into two paths, the particle will stay in path 1 in some of $x_4$, and stay in path 2 in the same portion of 
$x_4$, after pass the double slits, the two paths will interference each other, by equation (\ref{Superstition_1}), we have
\begin{equation}
|\triangle (x_4)^2| = |\frac{2\hbar}{p^{\alpha}} \triangle x_{\alpha}| |\psi_{path 1} +\psi_{path 2})|^2
\label{Superstition_2}
\end{equation}
At some spatial points (on the screen ),  $|\psi_{path 1} +\psi_{path 2})|$ becomes zero, at those points, equation
(\ref{Superstition_2}) could not be true for any non-zero $|\triangle (x_4)^2|$, it means that the density of $x_4$ 
is always zero at those points -- the particle does not go to those points. That explains the minima in interference fringes. 

Equation (\ref{Superstition_2}) also tells that at the same first dimension time t, X can be different value for different $x^4$, so the particle's position X is uncertain
in 4-dimensional time-space language, but it is unique in 3+3 dimensional time-space , i.e. at each 3-dimensional time $(x_0,x_4,x_5)$, particle P is always in 1 spatial location 
$X(x_1,x_2,x_3)$. We can imagine that, if we modify metric $g_{44}$ to the combination of different $\psi (p^\alpha_i)$, then particle can be at different momentums at the same time t. 

Why we always only get 1 ``dot'' at interference fringes for each photon ? Why 1 photon can not produce 2 ``dots'' on the screen? To produce a ``dot'' on the screen, the screen has to 
interact with photon, the interaction localized the photon and change photon's local curvature of time-space of photon. Let's say the interaction happened on 6-dimensional 
coordinate $A_0(x_0,x_1,x_2,x_3,x_4,x_5)$, which is spatial location $X_0(x_1,x_2,x_3)$ at 3-dimensional time $(x_0,x_4,x_5)$. After the interaction, the curvature
of local time-space of particle P changed, particle P can not move to another screen location $X_1(x_1,x_2,x_3)$ through $x_4$, the original geodesic of the particle is cut by
interaction. That corresponds to wave-packet collapse in quantum physics. 

There are two most important properties of 3-dimensional time: 

1) a particle can occupy more than 1 locations at the same 1st dimensional time t. 
2) Many particles can occupy the same location X at the same 1st dimensional time t. 

The first property gives us the non-local results of quantum physics and statistical interpretation of wave-function.
The second property will give us Bose-Einstein condensation. Considering two particles occupy location X at the same t, 
but their $(x_4,x_5)$ are always different at (X,t), then those two particles can not ``see'' each other, and they can not interact with each other ( unless
through 3rd particle). In a very small 3-D ball with many particles inside, if we can find a distribution of 2nd and 3rd time dimension of those particles such that 
$(x_4,x_5)_i <> (x_4,x_5)_j)$ always true at any time t and at any X (i, j are indices of particle), then those particles do not interact each other. 
If such distribution of $(x_4,x_5)_i$ exists, we get Bose-Einstein condensation for those particles. 
Does such distribution exist? If at time t, projection of the 6-dimensional geodesic to 3-D ball is always a loop (U(1) symmetry) at fixed time t, 
we can easily put many loops to SO(3) without cross each other, so all particles do not ``meet'' each other, we get Bose-Einstein condensation. 
But if the geodesic of the particle is not a loop, for example, if the geodesic of particle has SU(2) symmetry, 
as we know,  we can find a map from SU(2) to SO(3) which is double cover, i.e. one X only contains two different points of SU(2), 
suppose the two different points $(x_0,x_4,x_5)_i$, $(x_0,x_4,x_5)_j$  
from different particle, then  only 2 particles can put in a small space without meet with each other, 
this will explain why we have Pauli Exclusive Principle for SU(2) field, which only allows two electrons in the same ``location'', one with 
$\frac{1}{2}$ spin, the other one with $-\frac{1}{2}$ spin. 
We will talk about particle with $\frac{1}{2}$ spin later.

\section{Equations of 1-spin free particle} \label{OneSpin}

\subsection{Massless 1-spin free particle} \label{oneSpin_0}

The original 5-dimensional Kaluza field equations can be written as \cite{Overduin} :
\begin{eqnarray}
G_{\alpha\beta}                    = \frac{\kappa^2 \phi^2}{2} 
   T_{\alpha\beta}^{EM} - \frac{1}{\phi} \left[ \nabla_{\alpha} 
   (\partial_{\beta} \phi) - g_{\alpha\beta} \Box \phi \right] 
   \; \; \; , \nonumber \\
\nabla^{\alpha} \, F_{\alpha\beta} = -3 \, \frac{\partial^{\alpha} 
   \phi}{\phi} \, F_{\alpha\beta} \; \; \; , \; \; \;
\Box \phi = \frac{\kappa^2 \phi^3}{4} \, F_{\alpha\beta} 
   F^{\alpha\beta} \; \; \; 
\label{4dFieldEquns}
\end{eqnarray}
where $G_{\alpha\beta} \equiv R_{\alpha\beta} - R g_{\alpha\beta} / 2$ 
is the Einstein tensor,
$T_{\alpha\beta}^{EM} \equiv g_{\alpha\beta} F_{\gamma\delta} 
F^{\gamma\delta}/4 - F_{\alpha}^{\gamma} F_{\beta\gamma}$
is the electromagnetic energy-momentum tensor, and
$F_{\alpha\beta} \equiv \partial_{\alpha} A_{\beta} - 
\partial_{\beta} A_{\alpha}$.
Here the cylinder condition is being applied, which means dropping all derivatives 
with respect to the fifth coordinate.

For massless 1-spin free particle, if we write 6-dimensional metric as  
\begin{equation}
\left( \hat{g}_{AB} \right) = \left( \begin{array}{cc}
   g_{\alpha\beta} + A_{\alpha} A_{\beta} \; \; & \; \; 
      A_{\alpha} \;   \; \; \; \; \; \; \; \; \;   \\
   A_{\beta} \; \; & \; \; 
     1 \; \; \; \; \; \;  \\
    \; \; \;  & \; \;  \; \; \; \; \; \; \; \; \; \; \;   \;\;       -1  \end{array} \right) \; \; \; 
\label{6dMetric1S}
\end{equation}
and from the relation $\hat{g}_{\alpha\beta} \hat{g}^{\alpha\beta} = \delta^{\alpha}_{\beta} $, we have inverse metric:  
\begin{equation}
\left( \hat{g}^{AB} \right) = \left( \begin{array}{cc}
   g^{\alpha\beta}  \; \; & \; \; 
      - A^{\alpha} \;  \; \; \; \; \; \; \; \; \;   \\
   - A^{\beta} \; \; & \; \; 
     1+A^2 \; \; \; \; \; \;  \\
     \; \; \;  & \; \;  \; \; \; \; \; \; \; \; \; \; \;   \;\;        -1  \end{array} \right) \; \; \; 
\label{6dMetric1S_i}
\end{equation}
the $\alpha \beta-$, $\alpha 4- $, and 44-components of equations (\ref{5dEFE1}) become:
\begin{eqnarray}
G_{\alpha\beta} = \frac{1}{2} 
   T_{\alpha\beta}^{EM} \;, \; \;  
\nabla^{\alpha} \, F_{\alpha\beta} = 0 \; \; \; , \; \; \;
F_{\alpha\beta} F^{\alpha\beta} = 0 \; \; \; 
\label{6dFieldEquns1S}
\end{eqnarray}
The second of above equations is Maxwell equation. The third of equation is true for 
plane electromagnetic wave-function which 
is the case of single free-photon. If we ignore gravity (for free photon) in first item, 
Equation (\ref{6dFieldEquns1S}) is the equations for single 1-spin massless free particle. 

\subsection{1-spin free particle with mass $>0 $} \label{oneSpin_1}

For 1-spin particle with mass $m_0 > 0 $, we let 
\begin{equation}
\hat{A}_{\alpha} = A_{\alpha}e^{\frac{i }{\hbar} m_{0} x_5}
\label{hatA}
\end{equation}
where we choose $\hbar = 1$. and
\begin{equation}
\hat{A}_{5} = 0
\label{hatA_5}
\end{equation}
where $m_{0}$ is rest mass of particle, $x_5$ is 6th dimension coordinate. 
6-dimensional metric for 1-spin free particle become:
\begin{equation}
\left( \hat{g}_{AB} \right) = \left( \begin{array}{cc}
   g_{\alpha\beta} + \hat{A}_{\alpha} \hat{A}_{\beta} \; \; & \; \; 
       \hat{A}_{\alpha} \; \\
    \hat{A}_{\beta} \; \;                                & \; \; 
     1 \; \; \; \; \; \; \\
  \; \; \; \; \; \; \; \; \;  & \; \; \; \; \; \; \; \; \; \; \;  \; \;   -1  \end{array} \right) 
\label{6dMetric1S_m}
\end{equation}
Let $\hat{F}_{AB} \equiv \partial_{A} \hat{A}_{B} - 
\partial_{B} \hat{A}_{A}$, and A, B runs over 0,1,2,3,5 ( A, B $<>$ 4). 
Energy momentum tensor
\begin{equation}
\hat{T}_{AB} \equiv g_{AB} \hat{F}_{CD} \hat{F}^{CD}/4 - \hat{F}_{A}^{C} \hat{F}_{BD}
\label{1SpinT}
\end{equation}
where A,B run over (0,1,2,3,5, A, B $<>$ 4).
So the $\alpha \beta-$, $\alpha 4- $, and 44-components of 
6-dimensional Einstein equations (\ref{5dEFE1}) become:
\begin{eqnarray}
G_{\alpha\beta} = \frac{1}{2} 
   \hat{T}_{\alpha\beta} \; , \; \; \; 
\nabla^{\alpha} \, \hat{F}_{\alpha\beta} - m_0^{2} \hat{A}_{\beta} = 0 \; \nonumber \\
\frac{1}{4}\hat{F}_{\alpha\beta} \hat{F}^{\alpha\beta} - \frac{1}{2} m_{0}^{2} \hat{A}_{\alpha} \hat{A}^{\alpha} = 0 \; \; \; 
\label{6dFieldEquns1S_m}
\end{eqnarray}
where we choose $\hbar = 1$. This is equations for 1-spin single particle with mass $>$ 0. 
As we see above, the particle of 1-spin obtains its mass from derivative of 6th dimension (third time dimension).

\section{Equations of $\frac{1}{2}$-spin free particle} \label{halfSpin}

Equations for  $\frac{1}{2}$-spin free particles are Dirac equations:
\begin{equation}
(i\gamma^{\nu} \partial_{\nu} - m) \psi = 0
\label{DiracEquation}
\end{equation}
Here $\psi $ is four-component complex wave function, $\nu$ runs over 0,1,2,3. 
$\gamma ^{i}$, $i=0,1,2,3$ are $4\times 4$ complex
constant matrices, satisfying the relation 
\begin{equation}
\gamma ^{l}\gamma ^{k}+\gamma ^{k}\gamma ^{l}=2g^{kl}I,\qquad k,l=0,1,2,3.
\label{b1.2}
\end{equation}
where $I$ is the unit $4\times 4$ matrix, and $g^{kl}=$diag$\left(1,-1,-1,-1\right) $ is the metric tensor.

Dirac equations are not equations of single particle, it has 4-solutions corresponding 
to different states of electron. The wave-function of Dirac particle is:
\begin{equation}
\psi = \left(\begin{array}{cc}
	{\phi}_0 \\
	{\phi}_1 \\
	{\phi}_2 \\
	{\phi}_3 
   \end{array}\right)
\label{DiracWaveFun}
\end{equation}
where $\phi\alpha$ ($\alpha=0,1,2,3$) is 4-components of the 1st solution Dirac equation.

According to our hypotheses that different sates of particle
will have different local geometry metric, so the 4-solutions should have different metrics.
Let's start from first solution with spin $\frac{1}{2}$ and positive energy: 
 For $\psi $, in $x_3$ representation, Dirac equation can be re-write as:
\begin{equation}
\partial_{0} \phi_0 + \partial_{1} \phi_3 - i \partial_{2} \phi_3 + \partial_3 \phi_2 + im_0 \phi_0 = 0
\label{ModifiedDirac}
\end{equation}
Note: Start from this section, we will always choose $\hbar = 1 $.

Now for particle with half-integer spin, Let
\begin{eqnarray}
\hat{K_0} = Cg_{00}\phi_0 e^{im_0 x_5}   \nonumber \\ 
\hat{K_1} = Cg_{11}\phi_3 e^{im_0 x_5}  \nonumber \\ 
\hat{K_2} = -iCg_{22} \phi_3 e^{im_0 x_5} \nonumber \\ 
\hat{K_3} = Cg_{33}\phi_2 e^{im_0 x_5}  \nonumber \\ 
\hat{K_5} = Cg_{55}\phi_0 e^{im_0 x_5}   
\label{HalfSpinVector}
\end{eqnarray}
where $\phi_{\alpha} $ is the $\alpha$ components of equation (\ref{DiracWaveFun} ), C is 
constant to be determined, $g_{\alpha \alpha}$ is element of usual 4-dimensional 
metric (1,-1,-1.-1), $g_{55} = -1$.
$m_0$ is rest mass of the particle.

For single free particle with $\frac{1}{2}$-spin and positive energy, we choose metric as below:
\begin{equation}
\left( \hat{g}_{AB} \right) = \left( \begin{array}{cc}
   g_{\alpha\beta} + \hat{K}_{\alpha} \hat{K}_{\beta} \; \; & \; \; 
      \hat{K}_{\alpha} \; \; \; \; \; \;  \; \;  \hat{K}_{\alpha} \hat{K}_{5} \\
    \hat{K}_{\beta} \; \;  & \; \; 
     1 \; \; \; \; \; \;  \; \;   \hat{K}_5  \\ 
   \hat{K}_{5} \hat{K}_{\beta} \; \; & \; \;  \hat{K}_5 \; \; \; \;  \; \;  -1+\hat{K}_{5} \hat{K}_{5}  \end{array} \right) \;
\label{6dMetricHalfS}
\end{equation}
where $\alpha , \beta $ runs over 0,1,2,3, 
and 
\begin{equation}
\left( \hat{g}^{AB} \right) = \left( \begin{array}{cc}
   g^{\alpha\beta}  \; \;  & \; \; 
    - \hat{K}^{\alpha} \; \; \;  \; \; \; \\
   - \hat{K}^{\beta} \; \;  & \; \; 
     1+\hat{K}_{A}\hat{K}^{A} \; \; \; \;- \hat{K}^{5}\\
 \; \;  & \; \;  - \hat{K}^{5} \; \; \; \; \; \;   -1  \end{array} \right) 
\label{6dMetricHalfS_i}
\end{equation}

Let $\hat{E}_{AB} \equiv \partial_{A} \hat{K}_{B} - 
\partial_{B} \hat{K}_{A}$, and A, B runs over 0,1,2,3,5 ( A, B $<>$ 4). 
Define energy momentum tensor for half spin particle: 
\begin{equation}
\hat{T}_{AB} \equiv g_{AB} \hat{E}_{CD} \hat{E}^{CD}/4 - \hat{E}_{A}^{C} \hat{E}_{BD}
\label{halfSpinT}
\end{equation}
where A,B run over (0,1,2,3,5, A, B $<>$ 4),
so the AB-, A4-, and 44-components (A, B $<>$ 4) of 
6-dimensional Einstein equations (\ref{5dEFE1}) become:
\begin{eqnarray}
G_{AB} = \frac{1}{2} \hat{T}_{AB} \;, \; \; \; \;
\partial_{A} (\partial^{B} \hat{K}_{B}) - \partial^{B} \partial_{B} \hat{K}_{A} = 0   \nonumber \\
\frac{1}{4}\hat{E}_{AB} \hat{E}^{AB} = 0  
\label{halfspinEq}
\end{eqnarray}
To derive the equation above, we used the relation below:
\begin{equation}
\partial_{B} g^{BB} (\partial_{A} \hat{K}_{B} - \partial_{B} \hat{K}_{A} ) 
 = \partial_{A} (\partial^{B} \hat{K}_{B} ) - \partial^{B} \partial_{B} \hat{K}_{A} 
 \label{relation}
 \end{equation}

For free particle, it is reasonable to assume that each components of $\hat{K}$ satisfied plane-wave condition:
\begin{equation}
(-\partial^{\alpha} \partial_{\alpha} - m^2)\hat{K}_{\alpha}  = 0
\label{planewave}
\end{equation}
where $\alpha, \beta$ runs over 0,1,2,3, or equivalently:
\begin{equation}
\partial^{B} \partial_{B} \hat{K}_{A} = 0
\label{planewave_1}
\end{equation}
where A, B run over 0,1,2,3,5.
so the second equation of equations (\ref{halfspinEq}) become
\begin{equation}
\partial_{A} (\partial^{B} \hat{K}_{B})  =  0  \; \; \; \; for \, all \, A = 0,1,2,3,5
\label{halfSpinEq2}
\end{equation}
i.e. $\partial^{B} \hat{K}_{B} $ does not depended on $x_0,x_1,x_2,x_3,x_5$, so it is reasonable to let
$\partial^{B} \hat{K}_{B} $ equals zero. (Can not be a constant other than zero becaues $\hat{K}$ contains
a plane-wave function part of condition (\ref{planewave_1})).
Together with the third equation of 
equations (\ref{halfspinEq}), we have
\begin{eqnarray}
\partial^{B} \hat{K}_{B}  = 0  \;, \; \; \; \;
\hat{E}_{AB} \hat{E}^{AB} = 0 
\label{halfSpinEq3}
\end{eqnarray}
are the solutions of equation (\ref{halfspinEq}). Now we have 3 unknown functions: $\phi_0, \phi_2, \phi_3$ of
(\ref{HalfSpinVector}), we have above two equations, plus we need choose constant C to make the first 
equation of (\ref{halfspinEq}) becomes reasonable, and for $\phi_{\alpha}$, they should be normalized.
Substitute $\phi_{\alpha}$ into above $\hat{K}_A$, 
the first equation of (\ref{halfSpinEq3}) becomes Dirac equation (\ref{ModifiedDirac}), and the solutions 
of (\ref{halfSpinEq3}) are:
\begin{eqnarray}
\phi_0 = \sqrt{\frac{m_0+p_0 }{2m_0}}\;, \; \; \; 
\phi_1 = 0 \;, \; \; \; \nonumber \\
\phi_2 = \sqrt{\frac{m_0+p_0 }{2m_0}} \frac{p_3}{m_0+p_0} \;, \;  \; \; \;  \nonumber \\
\phi_3 = \sqrt{\frac{m_0+p_0 }{2m_0}} \frac{p_1 + ip_2}{m_0+p^0}  \; \; \; \; \; \; \\
\label{DComponents}
\end{eqnarray}
The above $\phi_{\alpha}$ is the solution of equations (\ref{halfspinEq}) and it is also first solution of 
Dirac equation (\ref{ModifiedDirac}). We choose the constant C in equations (\ref{HalfSpinVector}) as
\begin{equation}
C = \frac{\sqrt{(m_0+p_0)2m_0}}{p_3}
\label{Constant}
\end{equation}
Substitute (\ref{Constant}), (\ref{DComponents}) into (\ref{halfSpinT}),then:
\begin{equation}
\hat{T}_{ab} = p_{a}p_{b}e^{\frac{-2i}{\hbar}p^{c}x_{c}} 
\label{TAB}
\end{equation}
where $a,b,c$ runs over 0,1,2,3,5 and $p_5 = m_0$. The result is just we expected: the Energy-Momentum tensor becomes
the product of two 5-dimensional momentum vector $(p_{\alpha},m_0)$ times the square of plane wave function. 
Note: the above solutions is dervied under $x_3$ representation of Pauli
matrix, it means choose a special coordinate for our 6-dimensional metric. If we choose different representation,
the relationship between $\hat{K}_A$ and $\phi_{A}$ in (\ref{HalfSpinVector}) will be different, but equations 
(\ref{halfspinEq}), (\ref{halfSpinEq3}) does not dependent on the choice of coordinate, so we'll still get solutions
of Dirac equation under new representation.

If we let
\begin{eqnarray}
\hat{K_0} = Cg_{00}\phi_1 e^{ im_0 x_5}  \nonumber \\
\hat{K_1} = Cg_{11}\phi_2 e^{ im_0 x_5}  \nonumber \\
\hat{K_2} = iCg_{22}\phi_2 e^{im_0 x_5} \nonumber \\
\hat{K_3} = -Cg_{33}\phi_3 e^{im_0 x_5}  \nonumber \\
\hat{K_5} = Cg_{55}\phi_1 e^{im_0 x_5}  
\label{HalfSpinVector_2}
\end{eqnarray}
We can get local time-space metric of single free particle with spin $-\frac{1}{2}$, and the 
corresponding equation will derive the second solution of Dirac equation.
Similarly, we can get 
the local metric of single free particle with negative energy and spin $\frac{1}{2}$:
\begin{eqnarray}
\hat{K_0} = -Cg_{00}\phi_3 e^{im_0 x_5} \nonumber \\
\hat{K_1} = -Cg_{11}\phi_1 e^{im_0 x_5}   \nonumber \\ 
\hat{K_2} = iCg_{22} \phi_1 e^{im_0 x_5} \nonumber \\
\hat{K_3} = -Cg_{33}\phi_0 e^{im_0 x_5}  \nonumber \\ 
\hat{K_5} = Cg_{55}\phi_3 e^{im_0 x_5}  
\label{HalfSpinVector_3}
\end{eqnarray}
and the third solution of Dirac equation. And
\begin{eqnarray}
\hat{K_0} = -Cg_{00}\phi_4 e^{im_0 x_5}  \nonumber \\ 
\hat{K_1} = -Cg_{11}\phi_0 e^{im_0 x_5}  \nonumber \\
\hat{K_2} = -iCg_{22}\phi_0 e^{im_0 x_5} \nonumber \\
\hat{K_3} = Cg_{33}\phi_1 e^{im_0 x_5}  \nonumber \\
\hat{K_5} = Cg_{55}\phi_4 e^{im_0 x_5}  
\label{HalfSpinVector_4}
\end{eqnarray}
for single free particle with negative energy and spin $-\frac{1}{2}$,
and the 4th solution of Dirac equation.

As we see above, we didn't get all four Dirac solutions in one metric, instead, we get each solutions of
Dirac equation under each different metric, and each different metric corresponding to different state of 
particle. It is reasonable because: in this paper, the basic idea is that local metric of time-space
determined the state of particle, for a free single particle with one state (not combination of states), 
its local metric of time-space should only be one of solutions of Dirac equation. 
Thus, we can say that we obtained the single particle equations for the particle with $\frac{1}{2}$-spin in this section.
Note: the metric we got for $\frac{1}{2}$-spin particle is similar to the metric for 1-spin particle except that,
$\frac{1}{2}$-spin has nonzero $\hat{K}_5$ components. It causes the non-diagonal elements between 5th dimension
and 6th dimension. It indicates that for integer spin particle, the 5th dimension loops around 4-dimensional 
time-space, but for half integer spin particle, the 5th dimension loops around all other 5-dimensional time-space.
The metric of 1-spin particle has symmetry of 4-dimensional time-space, but
the metric of $\frac{1}{2}$ has symmetry of all 5-dimensions.

The similarity between the metric of single electron (\ref{6dMetricHalfS}) and the metric of a single photon (\ref{6dMetric1S})
makes us easier to combine eletron and photon into the same metric. If we use Klein's idea \cite{Kle26a}: 
the derivatives of its fifth coordinate not equal 0. We can interpret the non-zero derivatives of fifth coordinate as:
The coupling between electron and photon changed local time-space metric. 

Let
\begin{eqnarray}
\hat{K}_{A} \rightarrow \hat{K}_{A}e^{-i\gamma x_4}   
\label{4thDeriv}
\end{eqnarray}
where $\gamma $ is a very small constant, and there is no $\hat{K_4}$. Then
\begin{equation}
\partial_4 K_{A} = -i \gamma K_{A}
\label{partial5}
\end{equation}
The metric $\hat{g}_{AB} $ of the coupling of
electron-photon is:
\begin{equation}
\left( \begin{array}{cc}
   g_{\alpha\beta} + (A_{\alpha} A_{\beta} + \hat{K}_{\alpha} \hat{K}_{\beta}) \; \; & \; \; 
     A_{\alpha} + \hat{K}_{\alpha} \; \; \; \; \; \;  \hat{K}_{\alpha} \hat{K}_{5}  \\
    A_{\beta} + \hat{K}_{\beta} \; \; & \; \;
     1 \; \; \; \; \; \; \; \; \; \; \hat{K}_5 \\
   \hat{K}_{5} \hat{K}_{\beta} \; \; & \; \; \;\;  \hat{K}_5  \; \; \; \; \; \;  -1+ \hat{K}_{5} \hat{K}_{5}  \end{array} \right) 
\label{6dMetricCoupl}
\end{equation}
where $\alpha , \beta $ runs over 0,1,2,3. $A_{\alpha}$ is components of vector field of photon. $\hat{K}_{\alpha}$
satisfied one of equations (\ref{HalfSpinVector}), (\ref{HalfSpinVector_2}),(\ref{HalfSpinVector_3}),
(\ref{HalfSpinVector_4}). 
From above metric, one can obtain the coupling item: $-ieA_{\nu}\hat{K}_\nu $ and $\partial_{\alpha} \rightarrow \partial_{\alpha} - ie A_{\alpha} $. 
The metric of interaction between electron and photon will be discussed in future.

\section{Coupling of quantum field equations and gravity  } \label{gravity}

Kaluza's original purpose is to unify gravity and Maxwell equation. Since we already have above quantum field equations 
and they are all derived under 6-dimensional Einstein equations, if we assume that the gravity does not change the curvature 
of 5th and 6th dimension, it is straight forward to combine gravity and quantum field equations for particle with 0-spin, 
1-spin and $\frac{1}{2}$-spin.

For 0-spin particle, the metric is the same
\begin{equation}
\left( \hat{g}_{AB} \right) = \left( \begin{array}{cc}
   g_{\alpha\beta} \; \; & \; \;  \; \; \; \; \; \; \; \; 
       \\
   \; \; & \; \; 
      \phi^2  \; \; \;  \; \; \; \;  \; \;  \\
    \; \; & \; \; \; \;  \; \; \; \;  \; \; \; \;  \; \;   -1 \;
   \end{array} \right) \; \; \; ,
\label{5dMetric_0G}
\end{equation}
where  $g_{\alpha\beta}$ is 4-dimensional metric of original Einstein equation. The 5-dimension and 6-dimension 
are diagonal, so we can separate the contribution of gravity part and 5th and 6th dimension part. 
Field equations becomes:
\begin{eqnarray}
G^{E}_{\alpha\beta} + G^{Q}_{\alpha\beta} =  8\pi G T^{E}_{\alpha\beta} + T^{Q}_{\alpha\beta}  \; \; \; \\
\label{6dFieldEquns_g1}
G^{Q}_{5\alpha} = T^{Q}_{5\alpha} = -(\phi^{\star}\partial_{A}\phi) (\phi^{\star}\partial_{B}\phi) \\
\label{6dFieldEquns_g3}
\Box \phi - m_0^2\phi = 0  \; \; \;  \\
\label{6dFieldEquns_g2}
G^{Q}_{55} = T^{Q}_{55} = m_0^2
\label{6dFieldEquns_g4}
\end{eqnarray}
where G is gravational constant. $T^{E}_{\alpha\beta}$ is original Einstein 
energy momentum tensor,  $ T^{Q}_{AB}$ is quantum energy momentum tensor defined in section \ref{ZeroSpin}:
$T^{Q}_{AB} = p_{A}p_{B} $.
$G^{E}_{\alpha\beta}$ is original Einstein tensor, $G^{Q}_{AB}$ is quantum part of Einstein
tensor which is contributions from 5th and 6th dimension:
\begin{equation}
G^{Q}_{AB} \equiv R^{Q}_{AB} - R^{Q} \, \hat{g}_{AB} / 2
\end{equation}
and
\begin{equation}
R^{Q}_{AB} = -(\phi^{\star}\partial_{A}\phi) (\phi^{\star}\partial_{B}\phi)
\end{equation}
As we see that Klein-Golden equation becomes equation (\ref{6dFieldEquns_g2}).

For 1-spin particle with mass $>$ 0, the metric becomes 
\begin{equation}
\left( \hat{g}_{AB} \right) = \left( \begin{array}{cc}
   g_{\alpha\beta} + \kappa^2 A_{\alpha} A_{\beta} \; \; & \; \; 
      \kappa A_{\alpha} \;   \; \; \; \; \; \; \; \; \;   \\
    \kappa A_{\beta} \; \; & \; \; 
     1 \; \; \; \; \; \;  \\
    \; \; \;  & \; \;  \; \; \; \; \; \; \; \; \; \; \;   \;\;       -1  \end{array} \right) \; \; \; 
\label{6dMetric1S_mG}
\end{equation}
where $\kappa = 4\sqrt{\pi G} $. Field equations (\ref{6dFieldEquns1S_m}) should stay the same. 
If $m_0 = 0 $, the metric
above becomes original Kaluza metric with scalar field $\phi = 1 $.

For $\frac{1}{2}$-spin particle, the metric becomes:
\begin{widetext}
\begin{equation}
\left( \hat{g}_{AB} \right) = \left( \begin{array}{cc}
   g_{\alpha\beta} + \kappa^2 \hat{K}_{\alpha} \hat{K}_{\beta} \; \; & \; \; 
      \kappa \hat{K}_{\alpha} \; \; \; \; \; \;  \; \;   \kappa^2 \hat{K}_{\alpha} \hat{K}_{5} \\
    \kappa \hat{K}_{\beta} \; \;  & \; \; 
     1 \; \; \; \; \; \;  \; \;   \kappa \hat{K}_5  \\ 
    \kappa^2 \hat{K}_{5} \hat{K}_{\beta} \; \; & \; \;  \kappa \hat{K}_5 \; \; \; \;  \; \;  -1+\kappa^2\hat{K}_{5} \hat{K}_{5}  \end{array} \right) \;
\label{6dMetricHalfS_G}
\end{equation}
\end{widetext}
where $\kappa = 4\sqrt{\pi G} $. With gravity, we do not have simple relation as equation (\ref{relation}), since the derivative of 
$g^{\alpha \alpha} $ could be non-zero, we can not get free particle Dirac equation.
Instead, the equations below should still valid:
\begin{eqnarray}
G_{AB} = \frac{1}{2} \hat{T}_{AB} \;, \; \; \; \;
\nabla^{A} \, \hat{E}_{AB} = 0   \nonumber \\
\frac{1}{4}\hat{E}_{AB} \hat{E}^{AB} = 0  
\label{halfspinEq_g}
\end{eqnarray}
where A, B run over 0,1,2,3,5, the definition of $\hat{T}_{AB}$, $\hat{K}_{A}$, $\hat{E}_{AB}$ is defined the same as in section \ref{halfSpin}.

\section{Discussions and Conclusions} \label{SUM}

In the most part of this paper, we derived quantum field equations of 0-spin particle, massless 1-spin particle, 1-spin
particle with mass $>$ 0 and $\frac{1}{2}$-spin particles by using 6-dimensional metrics. The equations are derived naturally
as pure geometry properties of 6-dimensional time-space. Mass is included as derivative of 6th dimension. Metric of $\frac{1}{2}$-spin
particle has the same format as metric of 1-spin except that the 6th component of vector field is not zero. Particle's 
wave-function becomes geodesic of 6-dimensional time-space. As the conclusion, the field equations of those particles become
{\it pure geometry}, that makes us easier to unify gravity and quantum fields.

All of those can stay the same without interpreting the 5th and 6th dimensions are ``Time'' dimension. Why we need 
call them extra time dimensions ? To answer this question, we face with a much bigger question, what is time? It is too big 
question to answer here. Maybe we never get the answer. But we can show some critical properties of time. First,
Time makes us know the order of events happening. We know what happened before and what happened after. In the world 
1-dimensional time, we know one person can only do one thing at a time (depends on how do you define one thing). But
in quantum world, as we discussed in section III, quantum non-local effect makes a particle can show in different 
locations at the same time, the distance of those locations can be quite large as it is shown in Bell's inequality experiment.
That makes us to question that if there is more than 1 dimensional time in the world? Second, 
when we are talking about time in common life, we usually talk about ``proper time'' $\tau $, actually
time plays ``affine parameter'' in our time-space. In relativity, time is no longer as ``affine parameter'', time is ``affine 
parameter'' of geodesic only when we choose a special reference frame. As we see in section III, if we choose 5th dimension as parameter,
the particle's geodesic path is wave-function, it also naturally shows that the possibility to meet the particle is proportion to
square of wave-function. If no particle exists -- in vacuum, 5th dimension becomes affine parameter. That makes us believe 
that 5th dimension is time. Third, the extra dimensional time makes us
re-define the meaning of ``meet'' of two particles. In the new definition that two particle can meet each other if and only if
all 6 coordinates have the same value for both particle, i.e. their 6-dimensional geodesics have at least one point crossing each other.
That makes us to obtain the results of Bose-Einstein condensation. Actually the two properties of extra time dimension : 

1) a particle can occupy more than 1 locations at the same 1st dimensional time t. 
2) Many particles can occupy the same location X at the same 1st dimensional time t. 

They makes us to understand most of basic quantum phenomena. In addition, if we interpret 5th dimension as space, we have to
face the same problem as Kaluza: to make 5th dimension small. As we discussed in section III, we do not need make this assumption
as long as that in macrocosm world, the metric of 6-dimensional time ``localized'' all 3 time dimensions --- the collective effects
of enormous particles inside each objects in macrocosm world make all 3 time dimensions ``moves'' in the same behavior.

But why the 6th dimension is also time dimension. First, it is the consideration of symmetry. Since we have 3-dimensional space,
and we also need at least 2 dimensional time, and we need totally 6-dimensional time-space to derive all those equations,
why not the 6th dimension is time, which makes world 3D-Time + 3D-Space ? 
Second, as one notices that the metric is complex in this paper. That makes the definition of interval 
$ds$ becomes complex. 
The geodesic is complex function too and only related to 5th dimension. Actually
the metric in original Kaluza theory could be complex too since $A_{\alpha}$ is plane electromagnetic wave. 
But how do we understand complex interval? How do we understand complex geodesic ? To derive all the results in this paper
based on select a special coordinates. Can we choose different coordinate to make metric only containing real parts?
We know that a complex function can be described as a real function with two components. Is it 
possible that the reason we have complex metric is just because of trying to derive current quantum equations; 
it is possible that in future, we can make the metric which only contains real part by including second and third time dimensions
in geodesic? 

Now let's try to understand spin. As we see in 6-dimensional time-space metric of 0-spin particle, 5th dimension is diagonal. 
There is no components between
4-dimension and other dimensions. In time-space metric of 1-spin particle, 5th dimension wrapped around 4 dimensional time-space, no
components between 5th dimension and 6th dimension. 
In time-space metric of $\frac{1}{2}$-spin particle, the time-space geometry become more complicate, 5th dimension wrapped around 
all other 5 dimensional
time-space. In our knowledge of 4-dimensional world, if an objects moves around 
some space dimensions, it is a rotation movement. It is reasonable to interpret that the spin is the particle rotating the other 
dimensions through 5th dimension. 

As a conclusion, this paper shows us that we can describe quantum particle fields by using pure geometry methods. All we needs
are proper metrics and 6-dimensional Einstein field equations. The two hypotheses we used plus modified Kaluza 
metric are very good candidates to interpret quantum phenomena. The methods are simple in both logical and mathematical. 
We also demostrate the potentials to unify gravity
and other quantum fields by using just 6-dimensional time-space, since all those fields can be derived by Einstein equation. 
Finally, using pure geometry ---Einstein equation to describe quantumn particles is different from current gauge field theory, 
in section \ref{halfSpin}, we do not use all Dirac solutions (the 4-solutions), instead each time we only need one of Dirac solutions 
In other words, the methods we are using is focusing on interactions between single particles, it is particle theory rather than 
fields theory. We will discuss interactions between different particles in future.

\end{document}